\documentclass[10pt]{IEEEtran}
\usepackage{booktabs}

\usepackage{textcomp}

\usepackage{amsmath,amssymb}

\usepackage{amsthm}
\usepackage{amsfonts,mathrsfs}
\usepackage{color}
\usepackage{graphicx}
\usepackage{epsfig}
\usepackage{subcaption}
\usepackage[utf8]{inputenc}
\usepackage{amsmath}

\usepackage{enumitem}
\usepackage{accents}
\usepackage{mathtools} 
\usepackage{tabularx}
\newcolumntype{Y}{>{\centering\arraybackslash}X} 
\usepackage[prependcaption,colorinlistoftodos]{todonotes}
\usepackage{cancel}
\usepackage[normalem]{ulem} 
\usepackage{diagbox}

\usepackage{algorithm,algorithmicx}
\usepackage{algpseudocode}
\usepackage[hyphens]{url} 
\usepackage[hidelinks]{hyperref}

\definecolor{CBcyan}{rgb}{0.4,0.8,0.9333}    
\definecolor{CBred}{rgb}{0.9333,0.4,0.4667}  

\usepackage{tikz}
\usetikzlibrary{arrows.meta} 
\usetikzlibrary{fit, positioning}
\usetikzlibrary{calc}
\usetikzlibrary{decorations.pathmorphing} 

\tikzstyle{block} = [rectangle, minimum width=.7cm, minimum height=.7cm, text centered, draw=black]
\tikzstyle{tallblock} = [rectangle, minimum width=.5cm, minimum height=1cm, text centered, draw=black]
\tikzstyle{line} = [thick,-,>=stealth]
\tikzstyle{arrow} = [thick,->,>=stealth]
\tikzstyle{roundedblock} = [rectangle, minimum width=4cm, minimum height=2cm, text centered, draw=black, rounded corners=0.2cm]

\let\arg\relax
\DeclareMathOperator*{\arg}{\mathrm{arg}}

\newcommand{\bs}{\boldsymbol}
\newcommand{\mc}{\mathcal}
\newcommand{\bb}{\mathbb}
\newcommand{\R}{\bb R}

\DeclareMathAlphabet{\mathbbmsl}{U}{bbm}{m}{sl}

\DeclareMathOperator*{\argmin}{\operatorname{argmin}}

\newcommand{\col}{\operatorname{col}}

\newcommand{\Rmnum}[1]{\expandafter\@slowromancap\romannumeral #1@}

\newcommand{\smallsup}[1]{\text{\tiny \textnormal{#1}}} 

\usepackage{changepage}

\newcommand{\Cx}{D}
\newcommand{\Cu}{G}

\newcommand{\cx}{d}
\newcommand{\cu}{g}

\makeatletter
\newcommand{\specialcell}[1]{\ifmeasuring@#1\else\omit$\displaystyle#1$\ignorespaces\fi}
\makeatother
\newtheorem{assumption}{Assumption}
\newtheorem{proposition}{Proposition}

\newtheorem{definition}{Definition}
\newtheorem{lemma}{Lemma}

{\it}{}

\newtheorem{remark}{Remark}

\newcommand{\smallsum}{\mathop{\vcenter{\hbox{$\textstyle\sum$}}}\limits}
\newcommand{\tsum}{\textstyle\sum}

\newcommand{\VI}{\mathrm{VI}}

\newcommand{\AS}{\mc{A}}
\newcommand{\Cas}{C_{\AS}}
\newcommand{\Das}{D_{\AS}}
\newcommand{\cas}{c_{\AS}}
\newcommand{\lambdaas}{\lambda_{\AS}^*}
\newcommand{\Suas}{\mathcal{S}^{\text{u}}_{\AS}}
\newcommand{\Slambdaas}{\mathcal{S}^{\lambda}_{\AS}}

\newcommand{\nx}{n}
\let\nu\relax
\newcommand{\nu}{m}
\newcommand{\nc}{p} 

\begin{document}
\title{The explicit game-theoretic linear quadratic regulator for constrained multi-agent systems
}
\author{\IEEEauthorblockN{
        Emilio Benenati\IEEEauthorrefmark{1}, and
        Giuseppe Belgioioso\IEEEauthorrefmark{1}
    }\\
    \IEEEauthorblockA{
        \IEEEauthorrefmark{1}Division of Decision and Control Systems, KTH Royal Institute of Technology, Sweden}
    \thanks{Emails of the authors: \texttt{\{benenati, giubel\}@kth.se } }
    \thanks{This work was partially supported by the Wallenberg AI, Autonomous Systems and Software Program (WASP) funded by the Knut and Alice Wallenberg Foundation.} 
}
\maketitle

\begin{abstract}
We present an efficient algorithm to compute the explicit open-loop solution to both finite and infinite-horizon dynamic games subject to state and input constraints. Our approach relies on a multiparametric affine variational inequality characterization of the open-loop Nash equilibria and extends the classical explicit constrained LQR and MPC frameworks to multi-agent non-cooperative settings. A key practical implication is that linear-quadratic game-theoretic MPC becomes viable even at very high sampling rates for multi-agent systems of moderate size. Extensive numerical experiments demonstrate order-of-magnitude improvements in online computation time and solution accuracy compared with state-of-the-art game-theoretic solvers.
\end{abstract}

\section{Introduction}
Dynamic games model strategic interactions among self-interested rational decision makers whose actions influence and are influenced by the evolution of a shared dynamical system \cite{basar_dynamic_1999}. 
From a control perspective, dynamic games can be viewed as a multi-agent generalization of optimal control, in which each agent, the decision maker, faces an individual optimal control problem that is coupled to the others through the shared dynamics and the performance objective.

The most celebrated solution concept for dynamic games is the Nash equilibrium (NE), namely, a collection of control sequences (open-loop NE) or feedback laws (closed-loop NE) which are simultaneously optimal for all decision makers. In contrast to standard optimal control problems, dynamic games exhibit several distinctive  theoretical and computational challenges. For instance, the NE may be non-unique even in simple linear quadratic (LQ) settings with strongly convex objectives \cite{nortmann_feedback_2023, salizzoni_nash_2025}, and it can yield time-varying feedback laws that fail to stabilize the dynamics \cite{salizzoni_bridging_2025}. Moreover, available numerical methods are largely restricted to special classes of games. For LQ dynamic games, the NE can be computed by integrating the coupled state–costate dynamics on an invariant subspace \cite{engwerda_algorithms_2007, sassano_ol-ne_2025}. Another approach formulates the problem as a system of coupled algebraic Riccati equations, solved using iterative algorithms \cite{nortmann_nash_2024}. However, these techniques do not naturally extend beyond the unconstrained LQ setting.

In dynamic games subject to state and input constraints, the problem of computing a NE can instead be reformulated as a finite-dimensional variational inequality \cite{facchinei_finite-dimensional_2007} and solved using existing iterative algorithms \cite{di_local_2020}. While this approach does not directly extend to infinite-horizon settings, one can approximate the infinite-horizon NE solution by repeatedly solving a finite-horizon dynamic game and applying the first element of the resulting control sequence.
This control paradigm, known as Receding Horizon Games \cite{hall_stability_2024} or Game Theoretic Planning \cite{spica_real-time_2020} generalizes Model Predictive Control (MPC) to multi-agent non-cooperative settings, and has recently gained considerable research momentum in several research areas, including swarm robotics \cite{do_nascimento_game_2023}, energy management in smart grids \cite{hall_receding_2022}, smart mobility systems \cite{maljkovic_receding-horizon_2024}, autonomous driving \cite{fridovich-keil_efficient_2020, le_cleach_lucidgames_2021, peters_online_2023-1}, and racing, involving drones \cite{spica_real-time_2020}, various ground vehicles \cite{wang_game-theoretic_2021, kalaria_-racer_2025} and even Formula 1 cars \cite{fieni_game_2025}.s

Motivated by the growing interest in applications, several recent works have investigated theoretical performance and stability guarantees for game-theoretic MPC under various structural assumptions. For instance, \cite{hall_receding_2022, benenati_probabilistic_2024} establish closed-loop stability for dynamic games with a potential structure, while \cite{hall_stability_2024} extends these results to general-form dynamic games with stable dynamics. The work in \cite{hante_stabilizing_2025} proves closed-loop stability based on an approximate NE computation combined with a suitable terminal cost. Finally, \cite{benenati_linear-quadratic_2026} shows that infinite-horizon performance and stability can be achieved through an appropriate choice of terminal ingredients.

From a practical perspective, a major challenge lies in the computational burden of reliably solving a finite-horizon constrained dynamic game at each sampling time. To address this issue, several dedicated numerical solvers have been proposed, including methods based on Newton's type iterations \cite{lecleach_algames_2022, zhu_sequential_2023}, ADMM \cite{min_admm-iclqg_2025}, interior-point methods \cite{liu_log-domain_2025}, successive LQ approximations \cite{fridovich-keil_efficient_2020}, and active-set strategies \cite{laine_computation_2023}.

The practical applicability of these solvers in on-line settings is nevertheless limited by their convergence speed. For example, in an autonomous driving scenario with four vehicles, ALGAMES \cite{lecleach_algames_2022} and iLQGames \cite{fridovich-keil_efficient_2020}, two state-of-the-art solvers, report an average computation time of $860 \pm 251 \,\mathrm{ms}$ and $705 \pm 209 \,\mathrm{ms}$, respectively \cite[\S~5]{lecleach_algames_2022}. These computation times correspond to a control frequency of roughly $1 \,\mathrm{Hz}$, which is already borderline for autonomous driving and clearly inadequate for more time-critical applications such as autonomous racing.
In the latter case, trajectory-level planning typically operates at $5$–$20 \,\mathrm{Hz}$, while low-level vehicle dynamics are controlled at $20$–$100 \,\mathrm{Hz}$ \cite{kabzan_amz_2020, betz_tum_2023}.

An approach to mitigate this computational burden is proposed in \cite{liu_input--state_2024}, where the authors study the closed-loop stability of game-theoretic MPC under inexact computations, namely when only an approximate NE is available at each sampling instant. The limitations of this approach lie in the inherent trade-off between closed-loop performance and the available computational-time budget, as well as in the lack of analytical bounds on the iteration number to preserve nominal stability.

In this paper, we address this computational bottleneck in the LQ settings by developing an algorithm that constructs the explicit feedback policy of game-theoretic MPC, namely, the mapping from the initial state of the system to the resulting open-loop NE. This can be interpreted as an extension of the classical explicit MPC framework \cite{bemporad_explicit_2002} to multi-agent game-theoretic settings.
Our contributions are listed below.
\begin{enumerate}
\item[(i)] 
We derive an explicit characterization of the solution mapping of LQ dynamic games subject to state and input constraints as a piecewise affine (PWA) function of the initial state of the shared system. This directly yields an explicit PWA representation of the state-feedback policy in LQ game-theoretic MPC. Moreover, we extend this result to infinite-horizon dynamic games by leveraging their receding-horizon characterization derived in \cite{benenati_linear-quadratic_2026}.

\item[(ii)] We develop an efficient algorithm to compute this PWA state-feedback policy based on the state-space exploration technique proposed in \cite{arnstrom_high-performant_2024}. Unlike multi-parameter QP approaches, this method avoids the exponential branching typical of parameter-space partitioning, yielding improved scalability. To the best of our knowledge, this is the first algorithm to compute the explicit solution mapping of a constrained LQ dynamic game. 
\item[(iii)] We conduct extensive numerical studies to assess the limitations of the proposed approach and compare its performance against state-of-the-art solvers \cite{eckstein_operator-splitting_1998, min_admm-iclqg_2025, zhu_sequential_2023}. In scenarios with up to $8$ states and a $10$-step horizon, our approach achieves order-of-magnitude speedups in online computation time. Moreover, under a strict computational budget of $0.1,\mathrm{s}$, it yields a worst-case solution accuracy improvement of up to $10^{10}$ in terms of natural residuals. 
\end{enumerate}
 
Finally, we illustrate the real-time capabilities of the proposed approach on an autonomous driving scenario, showing that it enables the implementation of an overtaking maneuver while meeting a strict sampling rate requirement of 10 Hz.

\subsection{Notation and Preliminaries}
\subsubsection*{Basic notation}
For a matrix $M$, we define the weighted norm $\|x\|_{M}^{2} := x^\top M x$.  
For a collection of matrices (or vectors) $\{M_i\}_{i \in \mathcal{I}}$, we denote by 
$\col(M_i)_{i \in \mathcal{I}}$ their column-wise stacking.  
The symbol $\otimes$ denotes the Kronecker product.  
We write $\bs{0}_{m,n}$ and $\bs{1}_{m,n}$ for the $m \times n$ matrices of all zeros and all ones, respectively, and omit the dimensions when they are clear from context.  
Finally, we use $\mathbb{Z}_T := \{0,\ldots,T-1\}$ to denote the discrete-time index set of length $T$.

\subsubsection*{Operators theory}
We adopt standard
operator-theoretic notation and definitions from \cite{bauschke_convex_2017}. For a closed convex set $\mathcal{Y} \subseteq \R^n$, we denote the projection of $x \in \R^n$ onto $\mathcal{Y}$ by $\operatorname{proj}_{\mathcal{Y}}(x) = \argmin_{y \in \mathcal{Y}}~\|y-x\|$. The normal cone of $\mathcal{Y}$ is the operator $\operatorname{N}_{\mathcal{Y}} : \R^n \rightrightarrows \R^n$ defined by $\operatorname{N}_{\mathcal{Y}}(x) = \{v \in \R^n : \sup_{y \in \mathcal{Y}} ~ \langle v, y-x \rangle \leq 0\}$ if $x \in \mathcal{Y}$, and $\operatorname{N}_{\mathcal{Y}}(x) = \emptyset$ otherwise. We denote the graph of an operator $A: \R^n \rightrightarrows \R^n$ by $\operatorname{gra}_A = \{(x, y) \in \R^n \times \R^n : y \in A(x)\}$. We say that $A$ is ($\mu-$ strongly) monotone if $\langle x-y, u-v\rangle \geq 0 \, (\mu \|x-y \|^2)$ for any $(x, u) \in \operatorname{gra}_A$ and $(y, v) \in \operatorname{gra}_A$. For a single-valued operator $A: \mathbb{R}^n \rightarrow \mathbb{R}^n$ and a set $\mc B \subseteq \mathbb{R}^n$, the variational inequality $\VI(A, \mc B)$ is the problem of finding a vector $z^*\in\mc C$ such that $A(z^*)^\top(y-z^*) \geq 0$ for all $y\in\mc B$,
or, equivalently \cite[Eq. 1.1.3]{facchinei_finite-dimensional_2007}, such that $ 0\in\mc{N}_{\mc B}(z^*) + A(z^*)$. We denote as $\mathrm{SOL}(A, \mc B) := \left\{z \in \mathbb{R}^n ~|~ 0 \in  A(z) + \mc B(z) \right\} $ the set of solutions to $\VI(A, \mc B)$. If $A$ is affine and $\mc B$ polyhedral, then we refer to $\VI(\mc F, \mc C)$ as an affine variational inequality (AVI). 


\section{Problem formulation}\label{sec:prob_form}
We consider the multi-agent linear time-invariant system
\begin{equation} \label{eq:dynamics}
    x^{t+1} = A x^t + \sum_{i\in \mc I} B_i u_i^t,
\end{equation}
where $u^t_i\in\R^{\nu_i}$ is a control input determined by agent $i$, with $i \in \mc I := \{1,..., N\}$ denoting the set of agents, and $x^t\in\R^{\nx}$ is a shared state.
The goal of each agent $i\in \mc I$ is to minimize an individual quadratic performance criterion, defined over a control horizon  $\mathbb{Z}_T = \{0,1,\ldots,T-1\}$, namely,
\begin{align} \label{eq:objective}
        J_i(x, u_i) = \sum_{t=0}^{T-1} \frac{1}{2}\|x^t\|_{Q_i}^2  
        + \frac{1}{2}\|u_i^t\|_{R_i}^2,
\end{align}
with $u_i=(u_i^t)_{t\in \mathbb{Z}_T}$, $x=(x^t)_{t\in \mathbb{Z}_T}$, and $Q_i, R_{i}\succ 0$; while enforcing the following state and input constraints at all times:%
\begin{subequations}\label{eq:constraints}
    \begin{align} 
        \Cx x^t \leq \cx,  \label{eq:ol-NE:state_constr} \\
        \textstyle
        \Cu_i u_i^t +  \sum_{j\in\mc I\setminus\{i\}}\Cu_j u_j^t \leq \cu. 
        \label{eq:ol-NE:input_constr}
    \end{align}
\end{subequations}
Note that the performance criterion in \eqref{eq:objective} depends implicitly on the decisions of the other agents $u_{-i} := (u_j)_{j \in \mathcal{I}\setminus \{i\}}$, as the evolution of the shared state in \eqref{eq:dynamics} is determined by the control action of all agents $u:=(u_i)_{i\in\mc I}$.
%
Overall, each agent $i\in \mc I$ faces the following optimal control problem:%
\begin{subequations}
\label{eq:fin_hor_game}
\begin{align}\label{eq:fin_hor_game_cost}
\forall i \in \mc I: \quad 
\min_{u_i,\, \xi} & \quad  
\sum_{t\in \mathbb{Z}_T} 
\frac{1}{2}\|\xi^t\|_{Q_i}^2  
+\frac{1}{2}\|u_i^t\|_{R_{i}}^2 
\\
\text{s.t.}&\quad
\xi^{t+1} = A \xi^t + \sum_{j\in \mc I} B_j u_j^t,
\quad \forall t \in \mathbb{Z}_T
    \label{eq:constr_dyn} \\
&\quad
\Cx \xi^t \le \cx, 
\; 
\sum_{j\in \mathcal{I}} \Cu_j u_j^t \le \cu, 
\quad \forall t \in \mathbb{Z}_T
    \label{eq:constr_input_state} \\
    &\quad
    \xi^0 = x^0, \label{eq:constr_initial_state}
\end{align}
\end{subequations}
where $\xi$ is an auxiliary variable that tracks the shared state evolution, whereas $x^0$ is the actual initial state of system \eqref{eq:dynamics}.
The collection of these $N$ interdependent optimal control problems constitutes a \emph{dynamic game} with constraints.

We first focus on \textit{open-loop Nash equilibria} (ol-NEs) of the dynamic game \eqref{eq:fin_hor_game}. This solution concept assumes that the agents formulate their strategies at the moment the system starts to evolve, and cannot change them once the system runs. An ol-NE describes a strategy profile for which the agents are robust against any attempt by another agent to unilaterally alter their strategy. In other words, at an ol-NE, any agent that unilaterally changes their strategy will always be worse off.

Before formally introducing this solution concept, we write the game \eqref{eq:fin_hor_game} in a more compact form by using the dynamics \eqref{eq:constr_dyn} to eliminate the auxiliary state variable $\xi$, obtaining
\begin{subequations}\label{eq:game}
	\begin{align}
		\forall i \in \mathcal{I}: \quad
		\min_{u_i} & ~ f_i(u_i, u_{-i}, x^0) \\
		\text{s.t.} &~ (u_i, u_{-i})\in \mc U(x^0),
    \end{align}
\end{subequations}
where $f_i$ is a quadratic cost, parametrized in $x^0$, of the form 
\begin{align}
\label{eq:def_cost}
f_i(u,x^0) = \frac{1}{2} \|u_i\|^2_{H_{ii}} + u_i^\top F_i x^0 + \frac{1}{2}\|x^0\|^2_{Z_i}+ \sum_{j\neq i} u_i^{\top} H_{ij}u_j,
\end{align}
and $\mc U(x^0)$ is a polyhedral set of feasible input sequences, 
\begin{equation} \label{eq:rewritten_constraints}
	\mc U(x^0):= \{u \in \mathbb{R}^n ~|~ Cu + D x^0 \leq c\}.
\end{equation}
The constraints matrices $C,D,c$ and the quadratic cost functions $f_i$ are formally derived in Appendix \ref{app:AVI_derivation}.

\begin{definition}\label{def:ol-NE}
	A feasible collective input sequence $u^*\in\mc{U}(x^0)$ is an ol-NE at the initial state $x^0$ if, for each agent $i\in\mc I$:
	\begin{align}\label{eq:ol-NE}
				u^*_i \in \arg\min_{u_i}  ~~f_i(u_i, u_{-i}^*, x^0) \text{ s.t. } (u_i, u_{-i}^*)\in \mc{U}(x^0) 
	\end{align}
\end{definition}

Since the feasible set $\mc{U}(x^0)$  in \eqref{eq:rewritten_constraints} couples the decision of the agents, this definition corresponds to that of generalized Nash equilibrium (GNE) \cite[\S12]{palomar_convex_2010}. Therefore, for finite-horizon LQ dynamic games as in \eqref{eq:fin_hor_game}, the ol-NEs correspond to the GNEs of the compact game \eqref{eq:game}. Exploiting this equivalence, we can find one of such ol-NE by solving an affine variational inequality, parametrized in $x^0$, namely
%
\begin{align}\label{eq:VI}
\text{AVI}(\mathcal{F}(\cdot,x^0), \mc U(x^0)),
\end{align}
where $\mathcal{F}(\cdot,x^0) := (\nabla_{u_i}f_i(u, x^0))_{i \in\mathcal{I}}$ is the pseudo-gradient mapping of \eqref{eq:game}, obtained by stacking the partial gradients of the local functions $f_i$, and is an linear mapping of the form
\begin{equation}
\label{eq:PG}
\mathcal{F}(u,x^0) =	H u + F x^0.
\end{equation}
The matrices $H$ and $F$ are formally derived in Appendix \ref{app:AVI_derivation}.


Whenever $\mc U(x^0)\neq \varnothing$, the AVI in \eqref{eq:VI} admits a unique solution if the pseudogradient $\mathcal{F}(\cdot,x^0)$ is strongly monotone \cite[Prop. 2.3.3]{facchinei_finite-dimensional_2007}. Since $\mathcal{F}$ is a linear mapping, strong monotonicity for all parameters $x^0$ is equivalent to the matrix condition below \cite[Prop. 2.3.2]{facchinei_finite-dimensional_2007}, which we assume to hold true.%
\begin{assumption} \label{as:str_mon} 
The matrix $H$ in \eqref{eq:PG} is positive definite.
\end{assumption}

In the remainder of this paper, we focus on the offline computation of the parameter-to-solution mapping 
\begin{align}
	\label{eq:sol}
	\mathcal{S}: x^0 \mapsto \text{SOL}(\mathcal{F}(\cdot,x^0), \mc U(x^0 )),
\end{align}
over a compact set of initial conditions $\mathcal X \subset \mathbb{R}^n$. We refer to the problem of finding an explicit representation for $\mc S$ as a multiparametric AVI (mpAVI). 
The next sections presents two relevant scenarios where the mpAVI is of practical relevance.

\begin{remark}The results presented in the remainder of the paper can be readily extended to objective functions with coupling terms between agents' inputs of the kind $u_i^{\top}R_{ij} u_j$, with $j \neq i$. We omit these terms for ease of exposition.
\end{remark}

\subsection{Game Theoretic Model Predictive Control}
\begin{figure}
	\centering    \begin{tikzpicture}[scale=1.2, transform shape, baseline=(current bounding box.north), >=latex, node distance=1cm, auto]

        \node [draw, block] (plant) at (0,0) {{\footnotesize$x^+ = Ax + \smallsum_{i\in\mc I} B_i u_i$}};
        \node [draw, block, below=.2cm of plant] (GNEP) {{\footnotesize$\mathrm{SOL}(F(\cdot,x), \mathcal U(x))$}};
        \node [draw, block, left=.3cm of GNEP] (SelMat) {{\footnotesize$\Xi$}};

        \node [draw=none, left=.5cm of plant] (w) {};
        \node [fill=black, circle, right=.5cm of plant, inner sep=1pt] (dot) {};
        \node [draw=none, right=.5cm of dot] (x) {};

        \draw[-] (plant.east) -- node[above] {{\footnotesize$x$}} (dot.west);
        \draw[-] (dot.south) |- (GNEP.east);
        \draw[->] (dot.east) -- (x.west);
        \draw[->] (GNEP.west) -- (SelMat.east);
        \draw[->] (SelMat.north) |- node[above] {{\footnotesize$u$}} (plant.west);

    \end{tikzpicture}
	\caption{In game-theoretic MPC, control actions are generated by solving a finite-horizon dynamic game and applying the first element of the ol-NE control sequence to the system.}
	\label{fig:MPC}
\end{figure}
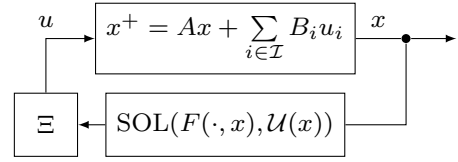

Game-theoretic MPC employs recursive ol-NE of finite-horizon dynamic games to approximate infinite-horizon solutions, while remaining computationally tractable and responsive to unforeseen disturbances. At each sampling time, the agents compute an ol-NE by solving the AVI in \eqref{eq:VI}, and then apply the first element of the ol-NE control sequence. This creates an implicit feedback policy $\kappa$, defined as
\begin{equation} \label{eq:control_law}
    \kappa(x) = \Xi \mathcal{S}(x), 
\end{equation}
where $\Xi$ is a selection matrix that extracts the first elements from the ol-NE sequence $\mathcal{S}(x)$. A schematic representation of this game-theoretic MPC controller is illustrated in Figure \ref{fig:MPC}.

Critically, implementing the control law in~\eqref{eq:control_law} requires computing the solution to the AVI in \eqref{eq:VI} within the desired sampling interval of the controlled system. Existing solvers for dynamic games are based on iterative algorithms and their convergence may be too slow for high sampling rate systems. Computing the explicit solution mapping $\mathcal S$ in \eqref{eq:sol} offline would enable on-line evaluation of ol-NEs with high accuracy without relying on costly iterative procedures.


\subsection{The Game-theoretic Linear Quadratic Regulator}

Ideally, one would solve the problem in \eqref{eq:fin_hor_game} over an infinite control horizon, namely, for $T\rightarrow \infty$, as that allows the agents to optimize over the entire trajectory of the system. Unfortunately, the problem of computing an ol-NE of an infinite-horizon games cannot be immediately recast to a finite-dimensional AVI, as that in \eqref{eq:VI}, due to the infinite number of decision variables and constraints. Nonetheless, under suitable assumptions on the game primitives, infinite-horizon ol-NEs can be recovered by solving a finite-horizon game equipped with appropriately chosen terminal ingredients \cite{benenati_linear-quadratic_2026}. Specifically, consider 
the fixed-point condition:
\begin{align}\begin{split}\label{eq:game_plus_term_cost}
		\forall i \in \mathcal{I}: \,
		u_i^*\in	\arg\min_{u_i} & ~ f_i(u_i, u^*_{-i}, x^0) + \varphi_i^{\text{term}}(u^*, x^0) \\
		\text{s.t.} &~ (u_i, u^*_{-i})\in \mc U(x^0),
	\end{split}
\end{align}
obtained by augmenting \eqref{eq:game} with the terminal cost
\begin{align}
\label{eq:terminal_cost}
\varphi_i^{\text{term}}(u^*, x^0):= \tfrac{1}{2} \|\xi^T\|^2_{P_i} + \big(S_i^\top x^{*,T} \big)^\top \xi^T,
\end{align}
where $P_i$ solves the algebraic Riccati equation
\begin{equation}\label{eq:standard_ARE}
	P_i = Q_i + A^{\top}P_iA - A^\top P_i B_i (R_i + B_i^{\top}P_iB_i)^{-1}B^\top P_i A,
\end{equation}
the matrix $S_i$ is given by $S_i = X_i - P_i$, with $(X_i)_{i\in\mc I}$ solving the systems of coupled matrix equations
\begin{equation}\label{eq:coupled_ARE}
	\forall i\in \mathcal{I}: \; \begin{cases} X_i = Q_i+A^{\top}X_i (A+B_i\tsum_{j\in\mc I}K_j),\\
		K_i = -R_i^{-1} B_i^\top X_i (A+B_i\tsum_{j\in\mc I}K_j),\end{cases}
\end{equation}
and $x^{*,T}$ is defined recursively as $x^{*,0} = x^0$, and $x^{*,t+1} = Ax^{*,t} + \sum_{i\in\mc I} B_i u^{*,t}_i$ for $t \in \mathbb{Z}_T$. This terminal cost is designed to reconstruct the contribution of the infinite-horizon cost-to-go truncated at time $T$.
Under some additional technical assumptions, Theorem 1 in \cite{benenati_linear-quadratic_2026} shows that $u_i^*$ which satisfies the fixed-point condition in \eqref{eq:game_plus_term_cost} coincides with the truncation of an infinite-horizon ol-NE.


Crucially, the fixed-point condition in \eqref{eq:game_plus_term_cost} does not correspond to that of a GNE, as $u_i^*$ appears in the argument of the optimization problem for agent $i$. Nevertheless, the fixed-point condition augmented with the terminal cost in \eqref{eq:terminal_cost} remains a set of coupled LQ optimization problems, and it therefore admits a parametric AVI reformulation akin to \eqref{eq:VI}. We refer the interested readers to \cite[Prop. 2]{benenati_linear-quadratic_2026} for the formal proof of this statement and a complete derivation of the VI formulation. Finally, we derive the multiparametric AVI corresponding to this infinite-horizon case in Appendix \ref{app:AVI_derivation} for completeness.

As a consequence, by constructing an explicit representation of the solution mapping of this AVI, we obtain an explicit state-feedback characterization of the infinite-horizon ol-NE.

\section{Multiparametric Affine VIs}\label{sec:solution}

For a fixed $x^0$, the AVI problem in \eqref{eq:VI} is equivalent to finding a primal-dual pair $(u^*, \lambda^*)$ that solves the KKT system%
\begin{subequations}
\label{eq:kkt}
\begin{align}
\label{eq:kkt-sta}
	0&= Hu + F x^0 + f +  C^\top \lambda \\
	0&\leq \lambda \; \bot - Cu - D x^0 + c \geq 0, 
	\label{eq:kkt-com}
\end{align}
\end{subequations}
where $\lambda$ is a dual variable associated with the constraints \eqref{eq:rewritten_constraints} \cite[Prop. 1.2.1]{facchinei_finite-dimensional_2007}.
If $\mc U(x^0)\neq \varnothing$, then \eqref{eq:kkt} admits a unique primal solution $u^*$ following the existence and uniqueness of the solution to \eqref{eq:VI}.
Denote as $\AS\subseteq\{1,\dots,\nc\}$ the set of all active constraints at $u^*$ and as $\Cas, \Das, \cas$ the rows of $C, D$ and $c$ (respectively) associated with $\AS$, that is,
 \begin{align}
 	\Cas u^* + \Das x^0 &=\cas. \label{eq:dual_active_constraints}
 \end{align}
Furthermore, denote as $\lambdaas$ the elements of $\lambda^*$ associated to the active constraints.  By the complementarity slackness condition in \eqref{eq:kkt-com}, the dual variables of the inactive constraints are $0$ and, thus, $\Cas^\top\lambdaas = C^\top \lambda^*$. Then, \eqref{eq:kkt} implies the linear system
 \begin{align}\label{eq:linear_system}\begin{split}
 		\begin{bmatrix} H & C^{\top}_{\AS} \\
 			C_{\AS} & \bs 0
 		\end{bmatrix}
 		\begin{bmatrix}
 			u^* \\ \lambdaas
 		\end{bmatrix} + \begin{bmatrix}F \\ D_\AS\end{bmatrix} x^0 + \begin{bmatrix} f \\ - c_\AS \end{bmatrix}  &=0.  
 	\end{split}
 \end{align}
 We assume for now that $\mc A$ is non-empty, and that $u^*$ satisfies the linearly independent constraint qualifications (LICQ)
 \begin{definition}
 	$u^*$ satisfies the LICQ if the rows of $C$ associated to its active set $\mc A$, namely, $\Cas$, are linearly independent.
 \end{definition}

Since $\Cas$ is full row rank and $H\succ 0$, the system \eqref{eq:linear_system}  admits a unique primal-dual solution \cite[Ex. 1.8.9]{facchinei_finite-dimensional_2007}
\begin{subequations}\label{eq:expression_of_solution}
\begin{align}
	u^*  &= - H^{-1}(F x^0 + f -  \Cas^\top \lambdaas)  \label{eq:primal_variable_expression} \\
	 \label{eq:dual_variable_expression}
	 \lambdaas &=-(\Cas H^{-1}\Cas^\top)^{-1}(\Cas H^{-1}(Fx^0+f) -\Das x^0 + \cas).
\end{align}
\end{subequations}
Now, let us define the primal and dual solution mappings
\begin{subequations}\label{eq:solution_mapping}
	\begin{align}
		\Suas:~& x^0 \mapsto \begin{cases}u^* ~ \text{as in} ~\eqref{eq:primal_variable_expression}  & \text{if}~\AS\neq\varnothing\\
			-H^{-1}(F x^0 + f)  & \text{if}~\AS=\varnothing
			\end{cases} \label{eq:solution_mapping:primal}\\
		\Slambdaas:~& x^0 \mapsto \col(\lambdaas, \bs{0}_{\nc-|\AS|}) ~ \text{with}~\lambdaas~\text{as in}~ \eqref{eq:dual_variable_expression} , \label{eq:solution_mapping:dual}
	\end{align}
\end{subequations}
which extend the explicit primal solution \eqref{eq:primal_variable_expression} to the case $\AS=\varnothing$, and pad the dual \eqref{eq:dual_variable_expression} with a vector of zeros associated with the inactive constraints. We will refer to $\mathcal{S}_{\mathcal{A}} = \Suas \times \Slambdaas$ as the $\AS$-active solution map. 

If $u^*$ does not satisfy the LICQ, then the matrix $\left[ \begin{smallmatrix}
	H & C^{\top}_{\AS} \\
	C_{\AS} & \bs 0
\end{smallmatrix} \right]$ is singular \cite[Ex. 1.8.9]{facchinei_finite-dimensional_2007}. The primal solution is still unique \cite[Prop. 2.3.3]{facchinei_finite-dimensional_2007}, and a primal-dual pair that solves \eqref{eq:kkt} -- and consequently solves \eqref{eq:linear_system} -- still exists \cite[Prop. 1.2.1]{facchinei_finite-dimensional_2007}. However, by the rank-nullity theorem, \eqref{eq:linear_system} admits infinite dual solutions. Furthermore, the matrix $(C_{\AS} H^{-1} C_{\AS} )$ is singular, thus the $\mc A$-active solution map in \eqref{eq:solution_mapping:dual} is not well-defined. In this case, however, we show as part of the proof of the following statement that a dual solution to \eqref{eq:linear_system} exists with zero elements associated to the linearly dependent constraints, and that the primal and non-zero elements of such dual solution are given by the $\bar{\AS}$-active solution, where $\bar{\AS}$ is a subset of $\AS$ with associated $C_{\bar{\mc A}}$ full row rank.
\begin{proposition} \label{prop:coverage_state_space}
	For each $x^0\in\mc X$, either of the following holds: 
	\begin{enumerate}
		\item[i)] $\mc U(x^0)=\varnothing$,
		\item[ii)] $\Suas(x^0)$ solves \eqref{eq:VI} for some $\AS\subseteq\{1,...,p\}$ such that $C_{\mc A}$ is full-row rank.
	\end{enumerate}
\end{proposition}
The proof is provided in Appendix \ref{app:proof:coverage_state_space}. The next proposition shows that the same $\mathcal{A}$-active solution mapping remains valid within a polyhedral region of the state space.
\begin{proposition}\label{prop:linear_mapping}
	Let $\AS\subseteq\{1,\dots, \nc\}$ such that $C_{\mc A}$ is full row rank. Then, $\mc S_{\mc A}(x^0)$ solves the KKTs \eqref{eq:kkt} for all $x^0$ such that%
	\begin{subequations} \label{eq:critical_region}
	\begin{align}
		C\Suas(x^0) + D x^0 &\leq c \\
		\Slambdaas(x^0)&\geq 0,
	\end{align}
\end{subequations}
\end{proposition}
The proof is given in Appendix \ref{app:proof:linear_mapping}. Since $\mc S_{\mc A}$ is affine, the inequalities \eqref{eq:critical_region} define a polyhedron in the space of $x^0$, which is commonly known as \emph{critical region} of the active set $\AS$. 
\begin{definition}\label{def:critical_region}
	For $\AS\subseteq\{1,\dots, \nc\}$ such that $C_{\mc A}$ is full row rank, its critical region is defined as
	\[
	\mc X_\AS := \{x^0\in\mc X ~|~ \eqref{eq:critical_region} ~\text{\emph{holds true}}\}.
	\]
	
\end{definition}

For a given $\mc A$, the corresponding critical region may be empty. Nevertheless, following Proposition \ref{prop:coverage_state_space}, the set of all critical regions covers the state space $\mc X$ for which the constraints in \eqref{eq:rewritten_constraints} are feasible. We denote as $q$ the number of non-empty critical regions. The AVI solution mapping $\mc S$ in \eqref{eq:control_law} can be explicitly formulated as
\begin{align} \label{eq:pwa_function}
	\begin{split}
		\mathcal{S}(x) = \begin{cases}
			\mc S^\smallsup{u}_{\AS_1}(x) & \text{if}~ x \in \mc X_{\AS_1}; \\ 
			\quad \vdots\\
			 \mc S^\smallsup{u}_{\AS_q}(x) & \text{if}~ x \in \mc X_{\AS_q}.
		\end{cases}
	\end{split}
\end{align}

The mapping \eqref{eq:pwa_function} is piecewise-affine (PWA), and can, in principle, be pre-computed offline by completing all possible combinations in $\{1,...,\nc\}$, where $\nc$ is the number of constraints. However, the number of combinations grows combinatorially with $p$, which in practice limits the applicability of this brute-force approach to small-scale problems. 
\begin{algorithm}[t]
\caption{Combinatorial active-set exploration for mpAVI}
\label{alg:mpVI_sol}
\begin{algorithmic}[1]
\Statex \textsc{Initialization}
\State Find $x^0$ such that $\mc U(x^0)\neq \varnothing$
\State Compute $u^*$ that solves \eqref{eq:kkt} \label{line:initial_solution}
\State $\AS \leftarrow \{i \in\{1,\dots,m\}~ |~C_i u^* + D_ix^0 + c_i = 0\}$ 
\State $\mc R \leftarrow \{\AS\}$, $\mc S \leftarrow \{\}$ 
\Statex \textsc{Main loop}
\While{$\mc{R}$ is not empty}
\State $\AS\leftarrow \text{FirstElement}(\mc{R})$
\State $\mc{R} \leftarrow \mc{R}\setminus \AS$
\If{$C_{\AS}$ full row rank \textbf{and} $\mc X_{\AS}\neq\varnothing$} 
\State Compute $\mc S_{\AS}$ as in \eqref{eq:solution_mapping}
\State $\mc S \leftarrow \mc S \cup \{(\mc S_{\AS}, \mc X_{\AS})\}$ 
\State $\mc R \leftarrow \mc R \cup\{\AS \cup \{i\}\}$, for all $i\notin\AS$
\State $\mc R \leftarrow \mc R \cup\{\AS \setminus \{i\}\}$, for all $i\in\AS$
\ElsIf{$C_{\AS}$ not full row rank \textbf{and}  $\mc X_{\AS}\neq\varnothing$}
\State $\mc R \leftarrow \mc R \cup\{\AS \setminus \{i\}\}$, for all $i\in\AS$
\EndIf
\EndWhile 
\State \Return $\mc S$
\end{algorithmic}
\end{algorithm}

\begin{figure}
\centering
	\includegraphics[width=0.85\columnwidth]{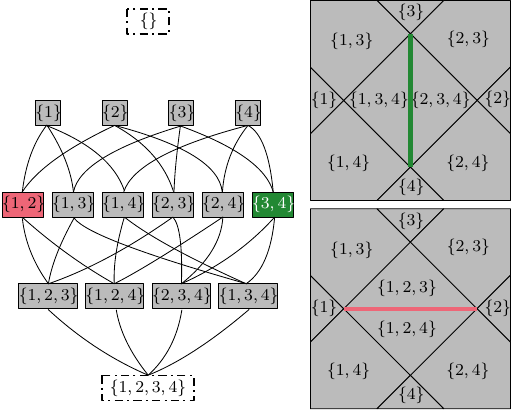}
	\caption{Left: combinatorial tree of constraint in \cite[Ex. 2]{ahmadi-moshkenani_combinatorial_2018}. Right: state space $\mc X = [-1.5,1.5]^2$ (drawn twice for clarity), partitioned in the critical regions found via Alg. \ref{alg:mpVI_sol}. The constraints corresponding to $\mc A=\{1,2,3,4\}$ are not linearly independent, thus the corresponding (non-empty) critical region is excluded from the state-space partition. For $\mc A=\{1,2\}$ and $\{3,4\}$ we find a one-dimensional critical region: the colored lines segment. Any path on the graph is a combinatorially valid sequence as for Definition \ref{def:comb_valid}. There is no edge leading to $\{\}$, as it exhibits an empty critical region. }
	\label{fig:example_crit_regions}
\end{figure}

\subsection{Efficient Construction of the Explicit Solution Mapping}
We explain how to reduce the burden of computing the explicit mapping \eqref{eq:pwa_function} by discarding combinations of constraints that are not instrumental in constructing a full-dimensional critical region. Following \cite{ahmadi-moshkenani_combinatorial_2018}, we construct a tree in which each node is a set of constraints $\mc A \subseteq \{1,...,\nc\}$, and its child nodes are all the sets of the form $\mc A \cup \{i\}$, $i\in\{1,...,p\}$.

For quadratic programs, the authors of  \cite{arnstrom_high-performant_2024} show that any two geometrically adjacent critical regions of the state space are connected in the tree and, crucially, the connecting path has some favourable structural properties that we formalize next, following \cite[Def. 5]{arnstrom_high-performant_2024} and \cite[Lemma 2]{arnstrom_high-performant_2024}.
\begin{definition} \label{def:comb_valid} A sequence of subsets $\AS^{(1)},\dots,\AS^{(n)} $ of $\{1,\dots,\nc\}$ is \emph{combinatorially valid} if, for all $k$:
	\begin{enumerate}
		\item[(i)] the critical region $\mc X_{\AS^{(k)}}$ is non-empty;
		\item[(ii)]  there exists $i\in\{1,\dots,\nc\}$ such that either \\ $\AS^{(k+1)} = \AS^{(k)} \cup \{i\}$ or $\AS^{(k+1)} = \AS^{(k)} \setminus \{i\}$;
		\item[(iii)]  either $C_{\AS^{(k)}}$ or $C_{\AS^{(k+1)}}$ is full-row rank.  
	\end{enumerate}
\end{definition} 

\begin{lemma}\label{le:combinatorial_connectivity} Consider any two $\AS$, $\AS'\subseteq\{1,...,\nc\}$ such that $C_\mc A$, $C_{\mc A'}$ are full-row rank, and their respective critical regions $\mc X_{\AS}$, $\mc X_{\AS'}$. Let $\mc X_{\AS}\cap\mc X_{\AS'}\neq \varnothing$. Then, there exists a combinatorially valid sequence that begins in $\AS$ and ends in $\AS'$.
\end{lemma}
The proof of Lemma \ref{le:combinatorial_connectivity} follows verbatim from that in \cite[Lemma 2]{arnstrom_high-performant_2024} and is thus omitted.
In words, Lemma \ref{le:combinatorial_connectivity} states that, starting from any non-empty critical region of a constraint set $\mc A$, one can reach any other by iteratively adding or removing a single constraint to $\mc A$, ignoring the constraint sets that result in empty critical regions. Furthermore, in view of item \emph{(iii)} in Definition \ref{def:comb_valid}, if by adding a constraint the full-rank condition is violated, then one should next remove a constraint, so that the condition is not violated for two successive active sets of the sequence.  This informs the development of Algorithm \ref{alg:mpVI_sol}. Note that its initialization requires a non-empty critical region (in Line \ref{line:initial_solution}). This can be computed  by evaluating the active set at the solution for a particular initial state via any standard VI solver, e.g. \cite{eckstein_operator-splitting_1998}. For every constraint combination, the offline computation complexity amounts to the solution of the linear system in \eqref{eq:expression_of_solution} and  of a feasibility problem. In Figure \ref{fig:example_crit_regions}, we illustrate an example of the critical regions computed via Algorithm \ref{alg:mpVI_sol}, and the corresponding tree of constraints set.

\begin{figure*}[t]
    \centering
    \begin{subfigure}{0.45\textwidth}
        \centering
        \includegraphics[width=.8\linewidth, trim=0 0 20 0, clip]{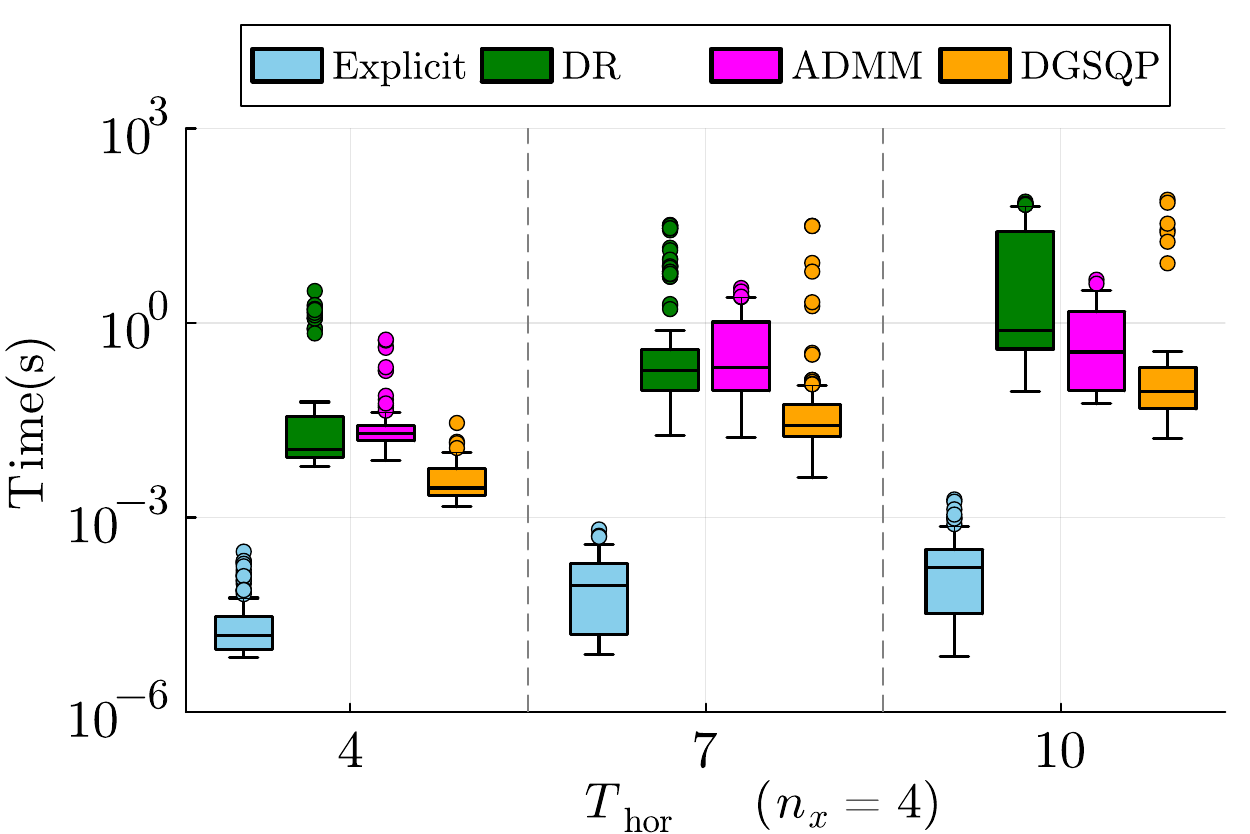}
    \end{subfigure}
    \begin{subfigure}{0.45\textwidth}
        \centering
        \includegraphics[width=.8\linewidth, trim=13 0 0 0, clip]{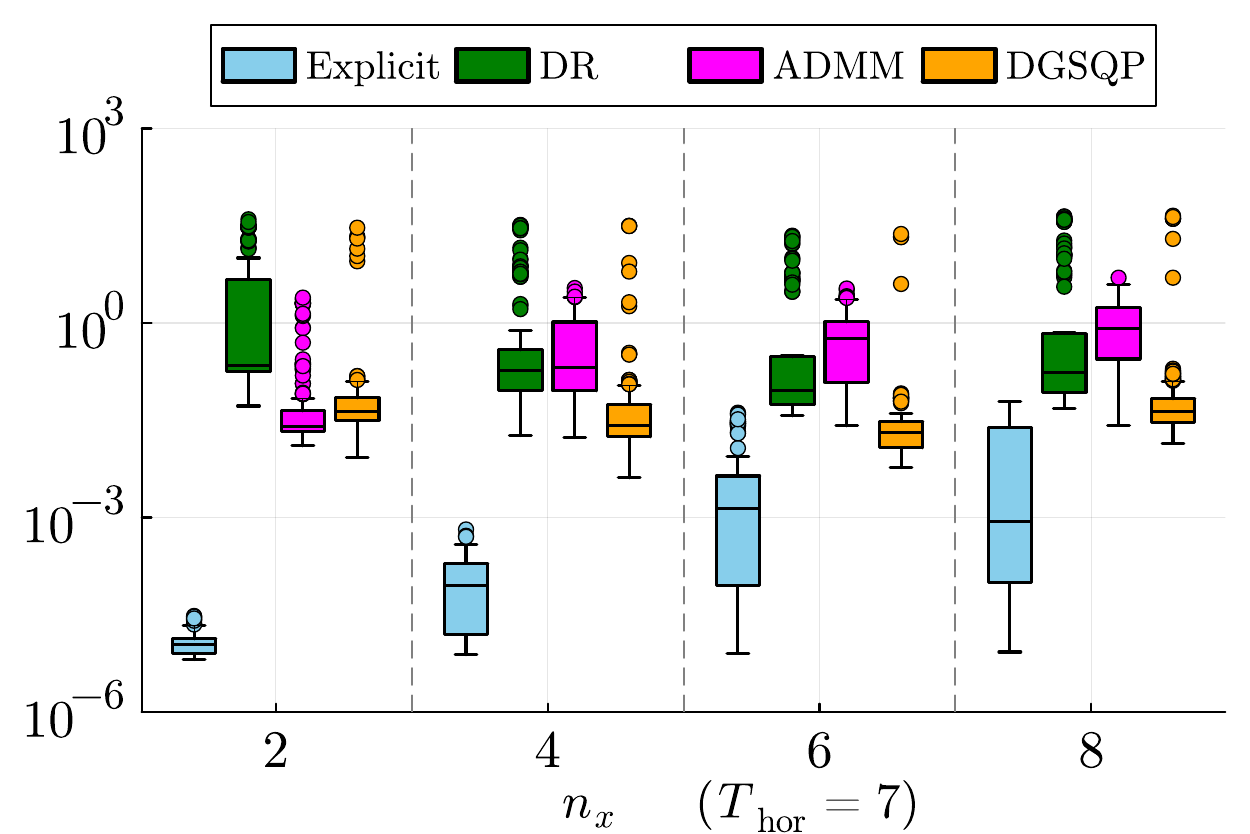}
    \end{subfigure}
    \caption{Online evaluation time of the explicit solution mapping \eqref{eq:pwa_function} constructed with Alg.~1, compared against the computation time for a high-quality solution ($r(u)=10^{-6}$) required by the Douglas--Rachford (DR) \cite{eckstein_operator-splitting_1998}, ADMM \cite{min_admm-iclqg_2025}, and DGSQP \cite{zhu_sequential_2023} solvers. The left plot shows the results in function of the horizon length, while the right one in function of the state dimension. \label{fig:OET}}
    \centering
    \begin{subfigure}{0.45\textwidth}
        \centering
        \includegraphics[width=.8\linewidth, trim=0 0 20 0, clip]{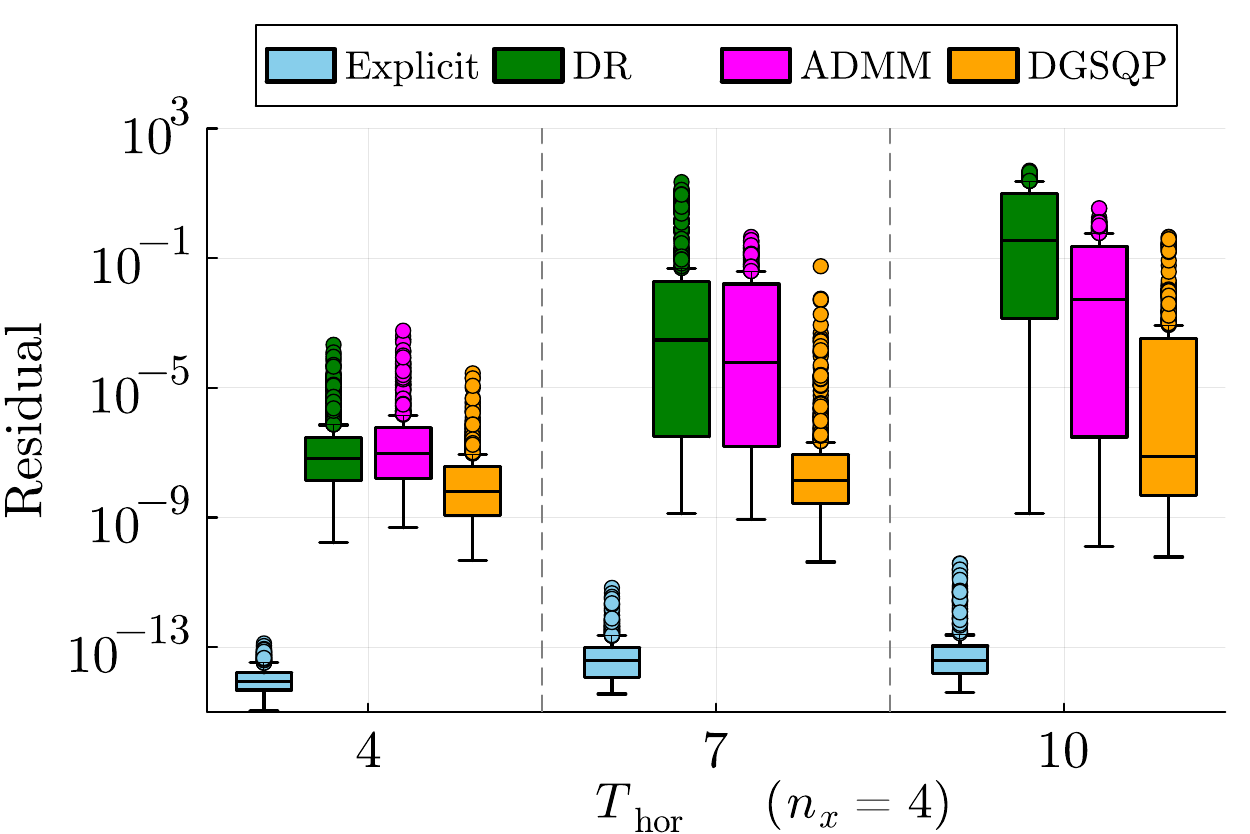}
    \end{subfigure}
    \begin{subfigure}{0.45\textwidth}
        \centering
        \includegraphics[width=.8\linewidth, trim=13 0 0 0, clip]{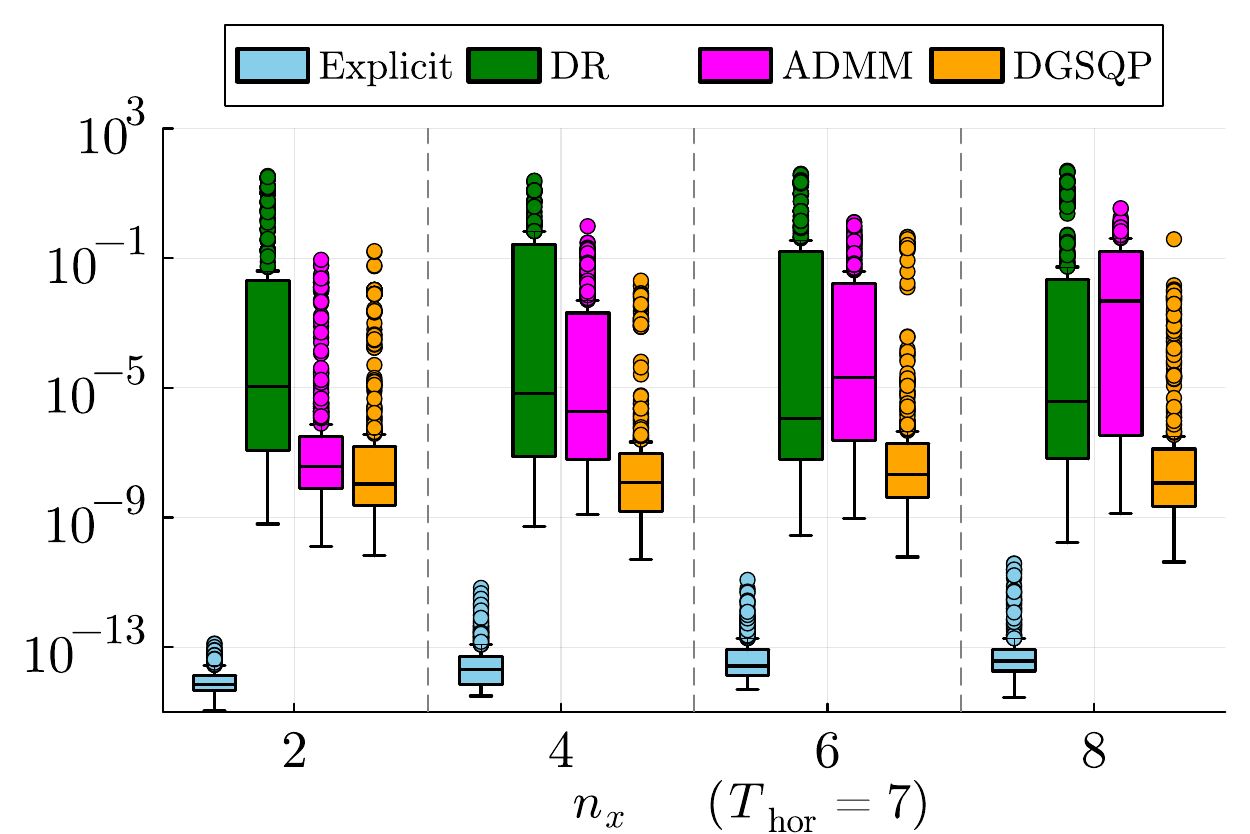}
    \end{subfigure}
    \caption{Solution quality, measured by the natural residual \cite[\S~6]{facchinei_finite-dimensional_2007}, of the explicit mapping computed with Alg.~\ref{alg:mpVI_sol}, compared with that obtained by the DR \cite{eckstein_operator-splitting_1998},  ADMM \cite{min_admm-iclqg_2025}, and DGSQP \cite{zhu_sequential_2023} solvers under a computational budget of $0.1\,\mathrm{s}$. The left plot reports the results as a function of the horizon length, and the right plot as a function of the state dimension.  \label{fig:accuracy}}
\end{figure*}
\section{Numerical results}
\subsection{Comparison against state-of-the-art iterative solvers}\label{sec:random_test}
We investigate the on-line and off-line computational performance of the proposed approach over a randomly-generated set of $100$ dynamic games, each of which is solved for $20$ initial states. The code used for generating the dataset and producing the results of this section is available online at \url{https://github.com/bemilio/scripts_for_explicit_LQGames_paper}. All simulations are implemented in Julia and run on a Dell Latitude 7450 with an Intel Ultra 7 165U chip and 32 GB RAM.

In our first experiment, we compare the on-line evaluation time of the explicit solution mappings built with Algorithm \ref{alg:mpVI_sol} against three state-of-the-art iterative solvers \cite{eckstein_operator-splitting_1998, min_admm-iclqg_2025, zhu_sequential_2023} set to compute a high-precision solution, measured in terms of the natural residual $r(u)=\|u - \mathrm{Proj}_{\mc U(x^0)}(u - \mc F(u, x^0))\|$. The iterations are terminated once  $r(u)<10^{-6}$, and the results are illustrated in Figure \ref{fig:OET}.
We observe that the online evaluation time of our approach grows sharper with the state dimension: its worst-case value rises from approximately  $10^{-4}\,\mathrm{s}$, for $n_x=2$, to $0.1\,\mathrm{s}$, for $n_x=8$. This is expected, as higher-dimensional problems generate more critical regions, making evaluation of the PWA mapping more demanding. Nevertheless, it remains faster than all solvers, the best of which, DGSQP \cite{zhu_sequential_2023}, exhibits worst-case times of roughly $10^2\,\mathrm{s}$ across all tested dimensions. Online evaluation time of the explicit mapping shows no substantial dependence on the horizon length, whereas iterative solvers are more affected.

In the second case study, we compare the solution accuracy of the proposed approach against the iterative solvers, under a computational budget of $0.1s$. The results, illustrated in Figure~\ref{fig:accuracy}, highlight a significant advantage of the explicit approach, as none of the tested solvers was able to compute reliably accurate solutions within the allotted time. This is to be expected, as our proposed method computes the NE exactly (up to machine precision) as the solution to a linear system.

\begin{table}[t]
	\begin{tabularx}{\linewidth}{|Y||Y|Y|Y|Y|}
		\hline
		\diagbox{$T$}{$n_x$} & \bfseries 2 & \bfseries 4 & \bfseries 6 & \bfseries 8 \\
		\hline\hline
		\bfseries 4  & {100\%} & {100\%} & {100\%} & {100\%} \\
		\hline
		\bfseries 7  & {100\%} & {100\%} & {100\%} & {66.7\%} \\
		\hline
		\bfseries 10 & {100\%} & {100\%} & {87.5\%} & {0\%} \\
		\hline
	\end{tabularx}
	\caption{Percentage of games whose solution mappings was computed within a 30 minutes runtime limit.}
	\label{tab:time_limit}
\end{table}
The main limitation of the proposed approach is the offline computational effort required to construct the explicit solution mapping. For each game instance, Algorithm~\ref{alg:mpVI_sol} is allocated a maximum of 30 minutes of offline computation time, after which it is terminated even if the mapping has not been fully constructed. In such cases, the algorithm returns a partially built mapping~$\mathcal{S}$ that does not cover the entire state space. Table~\ref{tab:time_limit} reports the percentage of instances for which the full solution mapping was entirely computed within the time limit. We observe that, for $n_x = 8$ and $T = 10$, the computation time limit was exceeded in all tested instances.

\begin{table*}
\centering
\renewcommand{\arraystretch}{1.25}
\begin{tabular}{l p{9cm} p{4cm}}
\toprule
\textbf{Maneuver phase} 
& \textbf{Cost functions of leading and following vehicles} 
& \textbf{Active safety constraint} \\
\midrule

\emph{Platooning} 
& 
$J_1 = \tfrac12 \sum_{t=0}^{T-1} 
\big( \|v_1^t - v_1^{\mathrm{ref}}\|^2 
+ \|l_1^t - l_1^{\mathrm{ref}}\|^2 
+ \|u_1^t\|_{R_{11}}^2 \big)$

\smallskip

$J_2 = \tfrac12 \sum_{t=0}^{T-1}
\big( \|p_2^t - p_1^t - d^{\mathrm{ref}}\|^2
+ \|v_2^t - v_2^{\mathrm{ref}}\|^2
+ \|l_2^t - l_2^{\mathrm{ref}}\|^2
+ \|u_2^t\|_{R_{2}}^2 \big)$
& 
$p_1^t - p_2^t \ge d_{\min}$ \\

\emph{Initiate overtake} 
&
Same $J_1$,  $J_2 = \tfrac12 \sum_{t=0}^{T-1} 
\big( \|v_2^t - v_2^{\mathrm{ref}}\|^2
+ \|l_2^t - l_2^{\mathrm{ref}}\|^2
+ \|u_2^t\|_{R_{2}}^2 \big)$,

\smallskip
with $l_2^{\mathrm{ref}}$ set to left-lane center
&
$p_2^t - p_1^t \le -\gamma(l_2^t - l_1^t) - d_{\min}$ \\

\emph{Perform overtake} 
& Same as \emph{initiate overtake}
& $l_1^t - l_2^t \ge d_{\min}$ \\

\emph{Complete overtake} 
& Same as \emph{initiate overtake}, with $l_2^{\mathrm{ref}}$ set to right-lane center
& $p_2^t - p_1^t \ge \gamma(l_2^t - l_1^t) + d_{\min}$ \\

\emph{Clear-road driving} 
& Same as \emph{complete overtake}
& none \\
\bottomrule
\end{tabular}
\caption{Summary of cost terms and safety constraints across the five overtake phases.}
\label{tab:overtake_cases}
\end{table*}

\subsection{Illustrative example: autonomous driving}
We demonstrate the proposed approach on an overtake maneuver involving two autonomous vehicles driving on a two-lane highway. Let $v_i$, $l_i$, $p_i$, $a_i$, and $\alpha_i$ denote respectively the speed, lateral position, longitudinal position, acceleration, and steering angle of vehicle $i\in\{1,2\}$. Each vehicle is modeled as a discretized unicycle \cite[Ex. 9.17]{sastry_nonlinear_2013} with sampling time $\Delta t = 0.1\,\mathrm{s}$. The input vectors and the joint state are, respectively, $ u_i = \col(a_i,\alpha_i)$ and $x = \col(v_1, l_1, p_2 - p_1,v_2,	l_2 )$.

\begin{figure}
	\centering
	\includegraphics[width=.9\columnwidth]{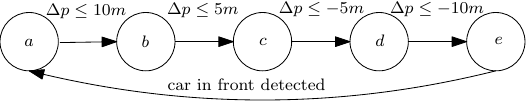}
	\caption{Finite-state machine that switches between the overtake controller cases. We denote $\Delta p = p_1-p_2$.}
	\label{fig:finite_state_machine}
\end{figure}

Depending on the relative positions of the two vehicles $\Delta p = p_1 - p_2$, we specify in Table \ref{tab:overtake_cases} the costs and constraints according to five phases of the overtake maneuvre: \emph{platooning}, \emph{initiate/perform/complete overtake}, and \emph{clear-road driving}. The finite-state machine in Fig.~\ref{fig:finite_state_machine} governs the transitions between these phases, using $\Delta p $ as the switching variable.

\begin{figure}
	\centering
\includegraphics[width=\columnwidth]{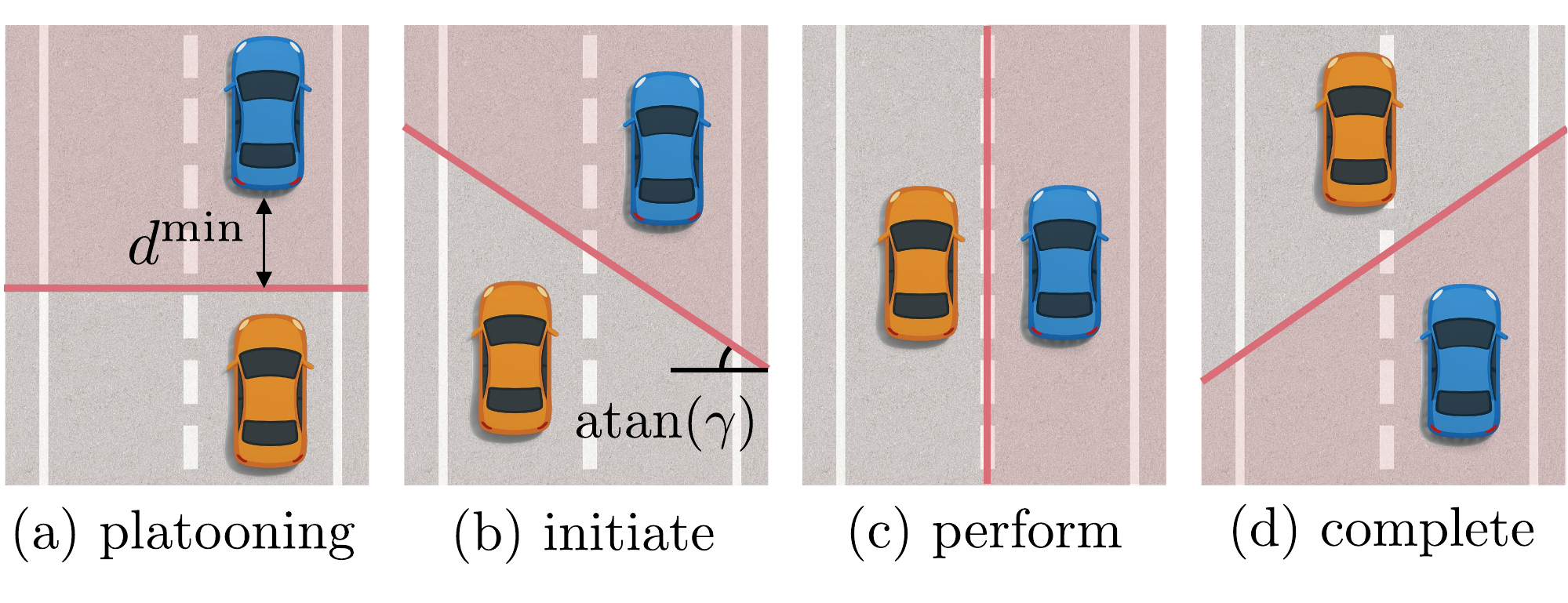}
	\caption{Safety distance constraint in the maneuvre phases.}
	\label{fig:distance_constraint}
\end{figure}

\paragraph{Platooning} The leading vehicle aims at maintaining a reference speed $v_1^{\mathrm{ref}}$ and lateral position $l_1^{\mathrm{ref}}$. The following vehicle aims at matching the leading vehicle's speed and maintaining a longitudinal distance $d^{\mathrm{ref}}$. 
The safety distance constraint $p_1 - p_2 \geq d^{\text{min}}$ is imposed to avoid collisions.
\paragraph{Initiate overtake} The following vehicle aims at reaching the overtake lane and the overtake reference speed, while satisfying the safety distance constraint $p_2-p_1 \leq - \gamma (l_2-l_1) - d^{\text{min}}$ at all times, where $\gamma>0$ determines the diagonal separation between vehicles, as illustrated in Figure \ref{fig:distance_constraint} (b). 
\paragraph{Perform overtake} Same costs as for the \emph{initiate overtake} phase, and with safety distance constraint $l_1-l_2\geq d_{\text{min}}$ to avoid lateral collisions between the vehicles.
\paragraph{Complete overtake} Same costs as for the \emph{initiate overtake} phase, except that the reference lateral position for the overtaking vehicle (2) is the right lane, and the safety distance constraint is $p_2-p_1 \geq \gamma (l_2-l_1) + d_{\text{min}}$.
\paragraph{Clear road driving} Same as for the \emph{complete overtake} phase, except that the safety distance constraint is removed. 

The explicit game-theoretic MPC controllers obtained by applying Algorithm~1 to the 5 maneuver phases is able to complete the entire overtaking maneuver successfully.\footnote{Animations available at:\\
\url{https://github.com/bemilio/scripts_for_explicit_LQGames_paper/blob/main/vehicle_animation.gif}} 
Throughout the maneuver, the maximum online evaluation time is $0.0122\,\mathrm{s}$, which is well below the sampling period required for $10$~Hz operation. This demonstrates that the explicit feedback law computed offline enables reliable real-time execution of a nontrivial multi-phase autonomous driving task.

\section{Conclusion}
We showed that the open-loop Nash equilibrium of a finite-horizon linear-quadratic dynamic game admits a piecewise affine dependence on the initial state. Leveraging this structure, we developed an algorithm that constructs the corresponding explicit solution mapping, effectively extending the paradigm of explicit MPC to non-cooperative multi-agent settings. As a result, linear-quadratic game-theoretic MPC becomes computationally feasible even at high sampling rates for multi-agent system of moderate size, as the explicit mapping removes the need for online iterative game-theoretic solvers.

\appendix
\subsection{Derivation of the AVI} \label{app:AVI_derivation}
We here report the derivation of $C,D,c, H_{ij},F_i, Z_i$ introduced in section \ref{sec:prob_form}. Denote:
\begin{align*}
	\Theta &:= \col(A^t)_{t\in\{1,...,T\}}, \\
	\Gamma_i &:= \begin{bmatrix}
		B_i & 0 & \dots & 0 \\
		A B_i & B_i & \dots & 0 \\
		\vdots &  & \ddots & \\
		A^{T-1}B_i & A^{T-2}B_i & \dots  &B_i
	\end{bmatrix}
\end{align*}
The dynamics in \eqref{eq:dynamics} is equivalently rewritten as
\begin{equation} \label{eq:dynamics_rewritten}
	x = \Theta x^0 + \left(\tsum_{i\in\mc I} \Gamma_i u_i \right).
\end{equation}
By stacking the constraints in \eqref{eq:constraints} and substituting \eqref{eq:dynamics_rewritten}, \eqref{eq:constraints} is equivalent to
\begin{align*}
	(I_T \otimes \Cx) \left( \Theta x^0 +  \tsum_{i\in\mc I}\Gamma_i u_i \right) & \leq \bs{1}_{T} \otimes \cx \\
	\tsum_{i\in\mc I} (I_T \otimes \Cu_i) u_i  &\leq \bs{1}_{T} \otimes \cu.
\end{align*}
From the latter, we find the expression for $C,D,c$ in \eqref{eq:rewritten_constraints}:
\begin{align*}
	C =&\begin{bmatrix}
		(I_T \otimes \Cx) \Gamma_1 & \dots & (I_T \otimes \Cx) \Gamma_N \\
		I_T \otimes \Cu_1 & \dots & I_T \otimes \Cu_N  
	\end{bmatrix}\\
	D= &\begin{bmatrix} (I_T \otimes \Cx) \Theta\\ 
		\bs{0} \end{bmatrix}\\
	c =& \begin{bmatrix}
		 \bs{1}_{T} \otimes \cx \\
		 \bs{1}_{T} \otimes \cu
	\end{bmatrix}.
\end{align*}
One can rewrite \eqref{eq:fin_hor_game_cost} as
\begin{align*}
	& \tfrac{1}{2}\xi^{\top} \bar{Q}_i \xi + \tfrac{1}{2} u_i^\top \bar{R}_{i} u_i,
\end{align*}
where we denoted
\begin{align}\label{eq:Q_bar}
&\bar{Q}_i= I_{T-1} \otimes Q_i ,\qquad \bar{R}_{i} = (I_T\otimes R_{i}).
\end{align}
By substituting \eqref{eq:dynamics_rewritten}, we obtain the following formulation for the matrices in \eqref{eq:def_cost}
 \begin{align}\label{eq:VI_matrices}
 	\begin{split}
 	H_{ii} &= \Gamma_i^\top \bar{Q}_i \Gamma_i + \bar{R}_{i},\\
 	H_{ij} &= \Gamma_i^\top \bar{Q}_i \Gamma_j, \qquad\qquad \forall j\neq i, \\
 	F_i &= \Gamma_i^\top \bar{Q}_i \Theta,\\
 	Z_i &= \Theta^\top\bar{Q}_i \Theta.
 	\end{split}
 \end{align}
 With the terminal cost in \eqref{eq:terminal_cost}, the problem in \eqref{eq:ol-NE} is cast as the AVI in \eqref{eq:VI} by substituting $\bar{Q}_i$ in \eqref{eq:Q_bar} with 
 \begin{equation}
 \begin{bmatrix}
 	I_{T-1} \otimes Q_i & 0 \\
 	0 & S_i + P_i
 \end{bmatrix}
\end{equation}
and then constructing $(H_{ij})_{i,j \in\mc I^2}$ and $(F_i)_{i\in\mc I}$ as in \eqref{eq:VI_matrices}.
\subsection{Proof to Proposition \ref{prop:coverage_state_space}}\label{app:proof:coverage_state_space}
 Let us consider a generic $x^0$ fixed such that $\mc U(x^0)\neq\varnothing$. Let $u^*$ be the unique primal solution to \eqref{eq:kkt},  $\AS$ the active set at $u^*$, $\Cas$ the associated rows of $C$, $\lambda^*$ a dual solution to \eqref{eq:kkt}, and $\lambda^*_{\AS}$ the elements of $\lambda^*$ with indexes in $\AS$.  We distinguish three cases: 
\paragraph{Case $\AS = \varnothing$} By the complementarity condition in \eqref{eq:kkt}, it must be $\lambda^*=\bs 0$. After substituting $\lambda^*=\bs 0$ in \eqref{eq:kkt}, it is immediate to see that  $\Suas(x^0)$ is the only primal solution.
\paragraph{$u^*$ satisfies the LICQ}
We showed in Section \ref{sec:solution} that, if $(u^*, \lambda^*)$ satisfy \eqref{eq:kkt}, then $(u^*, \lambda^*_{\mc A})$ satisfy \eqref{eq:linear_system}. $\mc S_{\mc A}(x^0)$ is a solution to \eqref{eq:linear_system}. By the uniqueness of the solution to \eqref{eq:linear_system} \cite[Ex. 1.8.9]{facchinei_finite-dimensional_2007}, it must be  $u^* =\Suas(x^0)$.
\paragraph{$u^*$ does not satisfy the LICQ}  We show that we can construct a set $\bar{\AS}\subset\AS$ of linearly independent constraints such that the $\bar{\AS}$-active solution map in \eqref{eq:solution_mapping} satisfies \eqref{eq:kkt}. Without loss of generality, assume $\lambda_{\AS}^*>0$. The argument applies unchanged after restricting $\AS$ to the indices corresponding to the nonzero elements of $\lambda^*$. Take $k\in\AS$, and scalars $(\alpha_j)_{j\in\mc A\setminus\{k\}}$ such that 
\begin{align} \label{eq:redundant_row}
	C_k = \tsum_{j\in\AS\setminus\{k\}} \alpha_j C_j,
\end{align}
where $C_j$ is the $j$-th row of $C$. Define the auxiliary multipliers
\begin{equation*}
	\lambda'_j:=\lambda^*_j + \alpha_j \lambda^*_k, \quad \forall j\in\AS\setminus\{k\}.
\end{equation*} 
We substitute \eqref{eq:redundant_row} in \eqref{eq:kkt} to get 
\begin{align}
	0&=Hu^* + F x^0 + f + C^\top \lambda^*  \Leftrightarrow \label{eq:kkt_primal_at_star}\\
	0&= Hu^* + F x^0 + f + \tsum_{j\in\AS} C_j^\top \lambda^*_j \Leftrightarrow \nonumber \\
	0 & =  Hu^* + F x^0 + f + \tsum_{j\in\AS \setminus\{k\}} C_j^\top (\lambda^*_j + \alpha_j \lambda^*_k ) \Leftrightarrow\nonumber\\
	0&=  Hu^* + F x^0 + f + \tsum_{j\in\AS \setminus\{k\}} C_j^\top \lambda'_j \Leftrightarrow \nonumber \\
	0&= Hu^* + F x^0 + f + C^\top \lambda', \label{eq:stationarity_new_multiplier}
\end{align}
where
$$\lambda' := \begin{bmatrix}
	\col(\lambda_j')_{j\in\mc A \setminus\{k\}}\\
	\bs 0_{\nc-|\AS|+1}
\end{bmatrix}.$$
If $\lambda'\geq 0$, then \eqref{eq:stationarity_new_multiplier} implies that $\lambda'$ is a dual solution with $|\mc A|-1$ non-zero elements. Otherwise, take
\begin{equation} \label{eq:definition_cvx_comb}
	\gamma = \max_{j ~\text{s.t.}~\lambda'_j<0}~~ \frac{\lambda'_j}{\lambda'_j - \lambda^*_j}.
\end{equation} 
Note that $\gamma\in(0,1)$. Construct the vector
\begin{equation}
	\lambda'' = (1-\gamma) \lambda' + \gamma \lambda^*. 
\end{equation}
For each $j$, following \eqref{eq:definition_cvx_comb} it holds that
\begin{equation*}
	\lambda''_j =  \lambda'_j + \gamma (\lambda^*_j-\lambda'_j) \geq  \lambda'_j - \tfrac{\lambda'_j}{\cancel{\lambda^*_j- \lambda'_j} } \cancel{(\lambda^*_j-\lambda'_j)}  \geq 0
\end{equation*}
and equality holds for the index that maximizes \eqref{eq:definition_cvx_comb}.	By taking a convex combination of \eqref{eq:kkt_primal_at_star} and \eqref{eq:stationarity_new_multiplier} with coefficient $\gamma$, the following holds true: 
\[ 0 = Hu^* + Fx^0 + f + C^\top (\gamma \lambda^* + (1-\gamma) \lambda'),
\] 
thus $\lambda''$ is a dual solution to \eqref{eq:kkt} with $|\mc A|-1$ non-zero elements. We repeat this procedure until we obtain a multiplier $\bar{\lambda}$ such that the  indexes of its non-zero elements  $\bar{\mc A}\subset\AS$ define a full-row rank matrix $C_{\bar{\AS}}$. Since $(u^*, \bar{\lambda})$ satisfy \eqref{eq:kkt}, then $(u^*, \bar{\lambda}_{\bar{\AS}})$ satisfy \eqref{eq:linear_system} for the index set $\bar{\AS}$. The unique solution to \eqref{eq:linear_system} is $\mc S_{\bar{\AS}}(x^0)$, thus it must be $\mc S_{\bar{\AS}}^\smallsup{u}(x^0) =u^*$. 
\subsection{Proof to Proposition \ref{prop:linear_mapping}} \label{app:proof:linear_mapping}
 We show that the $\AS$-active solution mapping computed at a generic $x^0$ that satisfies \eqref{eq:critical_region} also satisfies \eqref{eq:kkt}.
By definition, $\mc S(x^0)$ satisfies \eqref{eq:linear_system}. The primal condition in \eqref{eq:kkt} is immediately satisfied. Following \eqref{eq:critical_region}, the dual condition is satisfied if complementarity holds, that is,
\begin{equation}\label{eq:complementarity}
	(\Cas\Suas(x^0) + \Das x^0 -\cas)^\top \lambdaas =0.
\end{equation}
From the second row-block of \eqref{eq:linear_system}, we have
\begin{equation*}
	\Cas\Suas(x^0) + \Das x^0 -\cas=0,
\end{equation*} 
which implies \eqref{eq:complementarity} and concludes the proof.
\bibliography{bibliography}
\bibliographystyle{ieeetr}

\end{document}